# Spin system trajectory analysis under optimal control pulses

Ilya Kuprov

*School of Chemistry, University of Southampton,
Highfield Campus, Southampton, SO17 1BJ, UK*

Fax: +44 2380 594140

Email: i.kuprov@soton.ac.uk

## Abstract

Several methods are proposed for the analysis, visualization and interpretation of high-dimensional spin system trajectories produced by quantum mechanical simulations. It is noted that expectation values of specific observables in large spin systems often feature fast, complicated and hard-to-interpret time dynamics and suggested that populations of carefully selected *subspaces of states* are much easier to analyze and interpret. As an illustration of the utility of the proposed methods, it is demonstrated that the apparent "noisy" appearance of many optimal control pulses in NMR and EPR spectroscopy is an illusion – the underlying spin dynamics is shown to be smooth, orderly and very tightly controlled.

# 1. Introduction

Human brain, with its three-dimensional evolution history, often finds it difficult to visualize $2^N$-dimensional spin system trajectories, particularly for large values of $N$ and especially on Fridays. Yet such trajectories do occur in NMR pulse sequence analysis [1,2] and the problem is particularly severe for optimal control pulses [3-6] that feature complicated waveforms [5,6] that are not usually human-readable, either in their immediate shape or in the dynamics that they generate within the system.

In the context of magnetic resonance spectroscopy, the term "optimal control pulse" refers to a numerically optimized microwave or radiofrequency pulse designed to fulfill a set of difficult, but practically useful objectives, such as: ultrabroadband excitation at moderate power levels [5,7], highly selective excitation [8], resilience to $B_1$ field inhomogeneity [5,7], highly accurate coherence order transfer [9,10], calibration-free pulses [11] and so forth. Over the last ten years all of these objectives have been achieved with remarkable success – the community has seen pulses that excite 50 kHz bandwidth with 15 kHz RF power [11], beheld *JMR* [8], *JCP* [12] and *N.C.N.* [13] imprinted into spin excitation patterns of MRI samples and watched the magnetization being transferred with great accuracy across multi-spin chains [14]. Robust algorithms [4,15-18] and software [6,17,19] now exist for all those purposes.

Optimal control in magnetic resonance is a relatively simple special case of a much larger formalism [20,21] – for a spin system that must be steered from a state $\hat{\rho}(0)$ to a state $\hat{\sigma}$ in time $T$, the definition of transfer quality (known as *fidelity*) is:

$$f = \mathrm{Re}\langle\hat{\sigma}|\hat{\rho}(T)\rangle = \mathrm{Re}\left\langle\hat{\sigma}\left|\exp_{(O)}\left[-i\int_0^T\left(\hat{\hat{H}}(t)+i\hat{\hat{R}}\right)dt\right]\right|\hat{\rho}(0)\right\rangle \qquad (1)$$

where double hats denote superoperators, $\hat{\rho}(0)$ is the initial density matrix, $\hat{\sigma}$ is the density matrix of the desired transfer destination, $T$ is the experiment duration, $\exp_{(O)}$ indicates a time-ordered exponential, $\hat{\hat{R}}$ is a relaxation superoperator and $\langle\hat{a}|\hat{b}\rangle = \mathrm{Tr}(\hat{a}^\dagger\hat{b})$ is the scalar product in the density matrix space. The fidelity $f$ should be maximized as a functional of the parts of the Hamiltonian that can be experimentally controlled. Complete control over the system Hamiltonian is not usually available and it is formally split into two parts:

$$\hat{H}(t) = \hat{H}_0 + \sum_k c_k(t)\hat{C}_k \qquad (2)$$

where $\hat{H}_0$ is deemed beyond our direct influence and $\hat{C}_k$ are the operators whose amplitude our hardware can control – in the magnetic resonance context $\hat{C}_k$ are $\hat{L}_\text{X}$ and $\hat{L}_\text{Y}$ operators corresponding to radiofrequency or microwave fields. Their time-dependent coefficients $c_k(t)$ are usually discretized on a finite time grid and optimized as vectors [4,6,15,17].

Their remarkable performance notwithstanding, a noted feature of many optimal waveforms $c_k(t)$ is visual randomness (Figure 1 gives an example). Most researchers in the field have at some point been queried by a reviewer or a member of the audience as to why the supposedly optimal waveform "looks like noise". There is no denying that they often do [5,6,8,11,22], but we demonstrate below that this is an illusion – the underlying spin dynamics is very orderly. The demonstration of this fact required the development of visualization and similarity analysis methods for high-dimensional spin system trajectories: those methods are presented below, with optimal control NMR pulses used as illustrations.

## 2. Trajectory analysis strategies

As the right panel of Figure 1 demonstrates, simply plotting the amplitude and phase of each basis state as a function of time is not informative – fast oscillatory dynamics of specific states, in either time or frequency domain, is impossible to interpret directly. However, in our experiments with large-scale simulations [19] we found that populations of various physically relevant *subspaces* do have interpretable dynamics. In particular, the following classifications yield informative time dependence curves even with very high-dimensional trajectories:

1. <u>By populations of spin correlation orders</u>. In any direct product basis set, the correlation order of a state is defined as the number of non-unit spin operators in its direct product expression, for example:

$$\begin{array}{ll} \hat{\sigma}_\text{X} \otimes \hat{E} \otimes \hat{\sigma}_+ \otimes \hat{E} \otimes \hat{E} \otimes \hat{E} & k=2 \\ \hat{E} \otimes \hat{\sigma}_\text{Z} \otimes \hat{E} \otimes \hat{E} \otimes \hat{\sigma}_- \otimes \hat{\sigma}_\text{X} & k=3 \\ \hat{E} \otimes \hat{E} \otimes \hat{\sigma}_\text{Y} \otimes \hat{E} \otimes \hat{E} \otimes \hat{E} & k=1 \end{array} \qquad (3)$$

where $\hat{E}$ is the unit operator, $\hat{\sigma}_\text{XYZ}$ are Pauli matrices of appropriate dimension and $\hat{\sigma}_\pm$ are the corresponding raising and lowering operators. Because magnetic resonance simulations start and get detected in low correlation orders (1 in most cases and 2 for experiments involving singlet states), correlation order populations give a measure of complexity of a given trajectory. Classification of any state into correlation orders is always possible because the full state space $\mathfrak{L}$ of the spin system is a direct sum of correlation order subspaces $\mathfrak{L}_k$:

$$\mathfrak{L} = \mathfrak{L}_0 \oplus \mathfrak{L}_1 \oplus ... \oplus \mathfrak{L}_N \tag{4}$$

where $N$ is the number of spins in the system and $\mathfrak{L}_0$ only contains the unit operator. In any software implementation running in a direct product basis the population of a given correlation order $k$ in a state $\hat{\rho}$ is very straightforward:

$$p_k = \left\| \hat{\hat{P}}_{\mathfrak{L}_k} | \hat{\rho} \rangle \right\| \tag{5}$$

where $\hat{\hat{P}}_{\mathfrak{L}_k}$ is a projection superoperator into $\mathfrak{L}_k$. Because higher correlation orders relax faster [23] and are difficult to handle, a good control sequence would keep the population of high correlation orders low. An example of the improvement in readability brought about by Equation (5) is given in the middle panel of Figure 2 – after Equation (5) is applied, the system can be seen to move very smoothly from single-spin order subspace (where the initial state lives) into two- and three-spin orders, which then fade gradually to leave a single-spin order on the destination spin. This is in contrast to the complicated appearance of the control sequence that is shown in the left panel of the same figure.

2. <u>By population of spin coherence orders</u>. This is a generalization of the standard NMR coherence order diagrams [1,2] – in spherical tensor basis sets the coherence order of a state is defined as the sum of all projection quantum numbers in its direct product components, for example:

$$\begin{array}{ll} \hat{T}_{1,0} \otimes \hat{T}_{0,0} \otimes \hat{T}_{1,1} \otimes \hat{T}_{0,0} \otimes \hat{T}_{0,0} \otimes \hat{T}_{0,0} & m = 1 \\ \hat{T}_{0,0} \otimes \hat{T}_{2,2} \otimes \hat{T}_{0,0} \otimes \hat{T}_{0,0} \otimes \hat{T}_{1,-1} \otimes \hat{T}_{1,1} & m = 2 \\ \hat{T}_{0,0} \otimes \hat{T}_{0,0} \otimes \hat{T}_{2,0} \otimes \hat{T}_{0,0} \otimes \hat{T}_{0,0} \otimes \hat{T}_{0,0} & m = 0 \end{array} \tag{6}$$

where $\hat{T}_{l,m}$ are irreducible spherical tensor operators [24,25] and $\hat{T}_{0,0}$ is proportional to the unit matrix. Coherence orders also generate a partition of the full state space in a way similar to Equation (4):

$$\mathfrak{L} = \mathfrak{C}_{-M} \oplus \mathfrak{C}_{-M+1} \oplus ... \oplus \mathfrak{C}_{M-1} \oplus \mathfrak{C}_M \tag{7}$$

where $\mathfrak{C}_m$ is a subspace of all states with coherence order $m$. Coherence order may be negative and the maximum coherence order $M$ does not have to be equal to the number of spins in the system. For a given coherence order $m$ and a given state $\hat{\rho}$ the population is given by:

$$p_m = \left\| \hat{\hat{P}}_{\mathfrak{C}_m} | \hat{\rho} \rangle \right\| \tag{8}$$

where $\hat{\hat{P}}_{\mathfrak{C}_m}$ is a projection superoperator into $\mathfrak{C}_m$. Populations of coherence order subspaces give no indication of the complexity of dynamics (a state correlating the entire spin system can still have a zero coherence order), but they are useful in the analysis of liquid state NMR pulse sequences because the total projection quantum number remains invariant under liquid-state NMR drift Hamiltonians and provides a convenient illustration to the sequence mechanics [1,2]. Radiofrequency and microwave irradiation does, however, induce rotations between different coherence order subspaces and this classification is less useful in sequences involving continuous or closely spaced RF or MW events. An example of Equation (8) clarifying the dynamics under an optimal control sequence is given in Figure 4 – a complicated numerically optimized phase-modulated pulse is seen to be driving very smooth dynamics starting at zero-quantum coherence, moving through single-quantum coherences and into the destination, which is double-quantum coherence. Due to the high fidelity of the pulse (Figure 4, left panel), the destination state ends up being populated to very nearly its maximum possible amplitude ($1/\sqrt{2}$ in this case).

3. <u>By sum total of coherences and populations localized on each spin</u>. This is a further elaboration of Equation (4) that is often useful because the dynamics taking place in $\mathfrak{L}_1$ (the space of all single-spin populations and coherences) is particularly important. $\mathfrak{L}_1$ can be further partitioned into subspaces relating to individual spins:

$$\mathfrak{L}_1 = \mathfrak{L}_1^{(1)} \oplus \mathfrak{L}_1^{(2)} \oplus ... \oplus \mathfrak{L}_1^{(N)}, \qquad \mathfrak{L}_1^{(k)} = \mathrm{env}\left[ \mathfrak{su}(2s_k + 1) \right] \tag{9}$$

where the upper index in brackets enumerates spins, $2s_k + 1$ is the multiplicity of $k$-th spin and $N$ is the total number of spins in the system.

A significant obstacle to visualization is that spin dynamics in $\mathfrak{L}_1^{(k)}$ is often obscured by fast rotations caused by magnet and radiofrequency fields as well as quadratic interactions. We found that this problem disappears if the *total* *population* of each $\mathfrak{L}_1^{(k)}$ (which is of course invariant under unitary dynamics inside $\mathfrak{L}_1^{(k)}$) is considered:

$$p_k = \left\| \hat{\hat{P}}_{\mathfrak{L}_1^{(k)}} | \hat{\rho} \rangle \right\| \quad (10)$$

where $\hat{\hat{P}}_{\mathfrak{L}_1^{(k)}}$ is a projection superoperator into $\mathfrak{L}_1^{(k)}$. It should be noted that populations of two-spin subspaces may be evaluated in a similar way, but those subspaces are not in general a partitioning of $\mathfrak{L}_2$ because their intersections are not always empty. Equation (10) is useful in magnetization transfer experiments because it provides a measure of "total magnetization" (counting both populations and coherences) on each spin in the system. An example is given in the right panel of Figure 2 which reveals that the "noisy" optimal control pulse shown in the left panel is actually pushing the magnetization out of $C^{(\alpha)}$–H proton onto $C^{(\alpha)}$ carbon and from there onto C=O carbon in a very smooth and orderly way – something that would be quite contrary to intuition if only the pulse waveform were available for analysis.

4. <u>By sum total of coherences and populations involving each spin</u>. Equations (9) and (10) only include states that are *local* to a given spin. A complementary strategy is to examine population of the subspace spanned by all states that involve the current spin in any way, including correlations and coherences with other spins. For a given spin $k$, the system state space can be partitioned into:

$$\mathfrak{L} = \mathfrak{L}^{(k)} \oplus \left( \mathfrak{L} / \mathfrak{L}^{(k)} \right) \quad (11)$$

where $\mathfrak{L}^{(k)}$ is the subspace of all states that correlate that spin in any way and $\mathfrak{L} / \mathfrak{L}^{(k)}$ is the rest of $\mathfrak{L}$. The population of $\mathfrak{L}^{(k)}$ is then given by:

$$p_k = \left\| \hat{\hat{P}}_{\mathfrak{L}^{(k)}} | \hat{\rho} \rangle \right\| \quad (12)$$

where $\hat{\hat{P}}_{\mathfrak{L}^{(k)}}$ is a projection superoperator into $\mathfrak{L}^{(k)}$. This is a considerably broader definition than Equation (10) – it gives a measure of total involvement of a given spin at a particular stage of the pulse sequence. Consistently low involvement levels indicate that the spin can be dropped from the simulation altogether. To that end Equation (12) provides the benefit of a quantitative argument.

All four classification types suggested above are implemented in the trajectory analysis module of *Spinach* library [19] from version 1.2.1437 onwards and the simulations that generated Figures 1-5 are included in the example set that is supplied with the program.

## 3. Trajectory similarity scores

The other property that is hard to extract from the immediate appearance of either pulse shapes or system trajectories is the extent to which any two instances of system dynamics are "similar". Optimal control solutions are not unique – a different random initial guess in *e.g.* the GRAPE procedure [4,15] typically leads to a "different" pulse: the left panel of Figure 4 demonstrates complete lack of direct statistical correlation between two optimal control pulses that were obtained from different random initial guesses, but still accomplish the same goal (a transfer of magnetization between $^{1}H_{C\alpha}$ and $^{13}C_{O}$ in a protein backbone) with the same fidelity. A more sophisticated similarity criterion is therefore required for comparing given instances of spin system dynamics. From the algebraic perspective, two functions may be viewed as potentially useful:

1. <u>Running scalar product (RSP)</u>. A step-by-step scalar product between the corresponding vectors of the two trajectories:

$$s_{12}(t) = \langle \hat{\rho}_1(t) | \hat{\rho}_2(t) \rangle = \mathrm{Tr}\left(\hat{\rho}_1^\dagger(t)\hat{\rho}_2(t)\right) \qquad (13)$$

would return 1 if a pair of vectors is identical, $e^{i\varphi}$ if they are different by a phase, zero if they are orthogonal and the extent and phase of their overlap if they differ in a non-trivial way.

2. <u>Running difference norm (RDN)</u>. A step-by step norm of the difference between the corresponding vectors of the two trajectories:

$$d_{12}(t) = 1 - \frac{\|\hat{\rho}_1(t) - \hat{\rho}_2(t)\|}{2} = 1 - \frac{\sqrt{\langle \hat{\rho}_1^\dagger(t) - \hat{\rho}_2^\dagger(t) | \hat{\rho}_1(t) - \hat{\rho}_2(t) \rangle}}{2} \qquad (14)$$

would return 1 for identical vectors and zero if their tips are positioned on the opposite points of the unit ball that contains the trajectory. The choice of the norm rests with the user, but the Euclidean distance norm given in Equation (14) is likely the best choice in practice.

Both methods, however, are too sensitive in practice – a 90-degree difference in the phase of the magnetization vector makes the trajectories appear completely dissimilar on the RSP score ($\hat{L}_+$ is orthogonal to $\hat{L}_-$ in Liouville space) and very dissimilar on the RDN score, consistent with the large amount of liberty in the paths and phases that a system has between the source and the destination state – there are some points at which the two trajectories do not overlap at all on the RSP score. But the actual physical difference is minor – the magnetization passes through the same spin in a different phase. The definitions above should therefore be modified to reflect trajectory differences in a more informative way.

**State grouping (SG)**

The primary source of irrelevant phase differences is the rapid oscillation between $\hat{L}_X$ and $\hat{L}_Y$ operators under the offset part of the Zeeman Hamiltonian. These oscillations are easy to remove from visualization by considering *the total population of the subspace $\mathfrak{L}_\pm^{(k)}$ spanned by $\hat{L}_+^{(k)}$ and $\hat{L}_-^{(k)}$ operators of spin $k$* rather than their individual expectation values:

$$\left\{\left\langle\hat{L}_+^{(k)}\right\rangle,\left\langle\hat{L}_-^{(k)}\right\rangle\right\} \quad \rightarrow \quad \left\langle\mathfrak{L}_\pm^{(k)}\right\rangle = \sqrt{\left|\left\langle\hat{L}_+^{(k)}\right\rangle\right|^2 + \left|\left\langle\hat{L}_-^{(k)}\right\rangle\right|^2}, \quad \mathfrak{L}_\pm^{(k)} = \mathrm{span}\left\{\hat{L}_+^{(k)},\hat{L}_-^{(k)}\right\} \qquad (15)$$

The effect this transformation has on the similarity scores is illustrated in Figure 5 (center and right panels) – two different realizations of an optimal control trajectory moving the magnetization from $^1H_{C\alpha}$ to $^{13}C_O$ in a protein backbone fragment look very dissimilar, except for the initial and the final points, on both RSP and RDN scores (blue traces). However, state grouping using Equation (15) reveals that the difference is mostly in the phase of the magnetization vector – in other respects the trajectories are very similar (red traces, marked SG-RSP and SG-RDN respectively). It therefore appears, just as it did in the previous section, that the question of "*which subspaces does the system flow through?*" has a more interpretable answer than the same question about populations of individual states.

More generally, Equation (15) should be formulated in irreducible spherical tensor form:

$$\left\{\left\langle\hat{T}_{l,m}^{(k)}\right\rangle,\left\langle\hat{T}_{l,-m}^{(k)}\right\rangle\right\} \quad \rightarrow \quad \left\langle\mathfrak{T}_{l,\pm m}^{(k)}\right\rangle = \sqrt{\left|\left\langle\hat{T}_{l,m}^{(k)}\right\rangle\right|^2 + \left|\left\langle\hat{T}_{l,-m}^{(k)}\right\rangle\right|^2}, \quad \mathfrak{T}_{l,\pm m}^{(k)} = \mathrm{span}\left\{\hat{T}_{l,m}^{(k)},\hat{T}_{l,-m}^{(k)}\right\} \qquad (16)$$

where $\hat{T}_{l,m}^{(k)}$ is an irreducible spherical tensor operator with rank $l$ and projection $m$ on spin $k$. This formulation would also account for similar phenomena on spins greater than ½.

**Broad state grouping (BSG)**

Radiofrequency and microwave waveforms produced by optimal control methods typically cause rapid rotations within the entire $\{\hat{L}_X^{(k)}, \hat{L}_Y^{(k)}, \hat{L}_Z^{(k)}\}$ subspace of each spin. If the purpose of the visualization is to track magnetization transfer *between* spins, these rapid internal rotations are of no interest and may be removed altogether by extending Equations (15) and (16) to the entire state space of each individual spin. In the case of spin ½ we would have:

$$\{\langle \hat{L}_X^{(k)} \rangle, \langle \hat{L}_Y^{(k)} \rangle, \langle \hat{L}_Z^{(k)} \rangle\} \rightarrow$$
$$\langle \mathfrak{L}^{(k)} \rangle = \sqrt{|\langle \hat{L}_X^{(k)} \rangle|^2 + |\langle \hat{L}_Y^{(k)} \rangle|^2 + |\langle \hat{L}_Z^{(k)} \rangle|^2}, \quad \mathfrak{L}^{(k)} = \mathrm{span}\{\hat{L}_X^{(k)}, \hat{L}_Y^{(k)}, \hat{L}_Z^{(k)}\} \quad (17)$$

And in the case of arbitrary spin:

$$\langle \mathfrak{T}^{(k)} \rangle = \sqrt{\sum_{m=-l}^{l} |\langle \hat{T}_{l,m}^{(k)} \rangle|^2}, \quad \mathfrak{T}^{(k)} = \mathrm{span}\{\hat{T}_{l,-l}^{(k)}, \hat{T}_{l,-l+1}^{(k)}, \ldots, \hat{T}_{l,l}^{(k)}\} \quad (18)$$

This amounts to grouping the populations of the entire Lie sub-algebra of each individual spin – a map that may be schematically denoted as:

$$\mathfrak{su}(2s_1+1) \otimes \mathfrak{su}(2s_2+1) \otimes \ldots \otimes \mathfrak{su}(2s_N+1) \quad \rightarrow \quad \mathbb{R}^1 \otimes \mathbb{R}^1 \otimes \ldots \otimes \mathbb{R}^1 = \mathbb{R}^N \quad (19)$$

where $N$ is the number of spins in the system and $2s_k+1$ is the multiplicity of $k$-th spin. Equation (18) maps the population of each Lie algebra in the direct product into a one-dimensional subspace of a real vector space $\mathbb{R}^N$. Similarity scores computed for the trajectory image in $\mathbb{R}^N$ would only capture the *transfer* of coherence between spins – their internal dynamics would not be visualized.

When Equation (17) is used to group populations of closely related states, the two trajectories plotted in the right panel of Figure 4 turn out to be very similar – over 80% similarity throughout on RSP score and over 70% similarity on RDN score (green curves, labeled BSG-RSP and BSG-RDN respectively). This is in contrast to the complete lack of statistical correlation for the pulse shapes themselves (Figure 4, left panel).

## 4. Conclusions

We conclude that, for high-dimensional quantum trajectories, the visualization of subspaces that the spin system flows through is easier and more interpretable than the dynamics of individual observables or states. It was demonstrated above that those subspaces may be tailored to specifically monitor the dynamics of interest and remove less relevant information from the picture. Several specific classes of subspaces are offered to that end.

The resulting visualization methods revealed that the noisy appearance of optimal control pulses is an illusion – throughout the example set we see orderly transitions from the initial condition into appropriate correlations with relevant spins to exactly the right level so as to execute the required transfer with the highest possible accuracy. A fitting analogy here would be with the coordinates of a shepherd dog steering a herd of sheep. Taken separately, its behavior would seem chaotic – yet it exerts very precise control and eventually gets the herd to a designated location without losing a single sheep on the way.

## Acknowledgements

The author is grateful to Christiane Koch, Malcolm Levitt, Burkhard Luy, Niels Chr. Nielsen, and Konstantin Pervushin for useful discussions. This work is supported by EPSRC (EP/H003789/1) and FP7 (297861/QUAINT).

**Figure captions**

**Figure 1**  An illustration to the fact that most optimal control pulse waveforms are not directly interpretable. **Left panel:** phase profile of a phase-modulated broadband excitation pulse that meets the following requirements: $\hat{L}_\text{Z} \rightarrow \hat{L}_\text{X}$ excitation with at least 99% fidelity for a 50 kHz frequency range; constant RF power level of 15 kHz; tolerance for $B_1$ inhomogeneity of ±30%; pulse duration 1.0 ms; 625 time discretization points. See Ref. [5] for further information on such pulses. **Right panel:** Bloch sphere representation of the dynamics of a spin that is off resonance by 250 Hz under the pulse described above. The spin eventually arrives onto the X axis with the prescribed fidelity, but its intermediate dynamics is obscure.

**Figure 2**  Analysis of spin system dynamics under an optimal control pulse designed to move all magnetization from the $C^{(\alpha)}$–H proton of a protein backbone fragment (Figure 3) to the C=O carbon without leaking any magnetization to other nearby spins. **Left panel:** control operator coefficients (fractions of the nominal power level) as functions of time for the optimal solution (99% transfer fidelity). **Middle panel:** spin system dynamics, classified into spin correlation orders using Equation (5). **Right panel:** further analysis of the dynamics in the single-spin order subspace using Equation (10) – note the orderly transition from $C^{(\alpha)}$–H proton, over to $C^{(\alpha)}$ carbon and onwards to the C=O carbon, in contrast with the noisy appearance of the numerically optimized pulse that is driving the system.

**Figure 3**  Relevant interaction parameters of the protein backbone fragment used to generate the pulse given in Figure 2. The simulation in question is a part of the example set supplied with our *Spinach* spin dynamics simulation library [19]. The magnetic induction is set to 9.4 Tesla.

**Figure 4**  Analysis of spin system dynamics under an optimal control pulse designed to move the population of the $\hat{T}_{1,0}$ state of a spin-1 particle with a rhombic quadrupolar interaction to the $\hat{T}_{2,2}$ state to the maximum possible extent (70.7% is the Sorensen bound in this case). **Left panel:** phase profile of the numerically optimized pulse. **Right panel:** spin system dynamics, classified into coherence orders according to Equation (8). Single-quantum coherence can be seen

accumulating and then fading in a very precise sequence as the system climbs into the double-quantum coherence under the influence of the control sequence.

**Figure 5** Trajectory similarity analysis for two optimal control pulses solving the same state transfer problem (described in the caption to Figure 2) to the same fidelity, but obtained from different random initial guesses. **Left panel:** a demonstration of the lack of direct statistical correlation between the two solutions. **Middle panel:** running scalar product similarity score for the two system trajectories without preprocessing (RSP, blue curve), with similar states grouped using Equation (15) (SG-RSP, red curve) and with similar states grouped using Equation (17) (BSG-RSP, green curve). **Right panel:** running difference norm similarity score for the two system trajectories without preprocessing (RDN, blue curve), with similar states grouped using Equation (15) (SG-RDN, red curve) and with similar states grouped using Equation (17) (BSG-RDN, green curve).

FIGURE 1

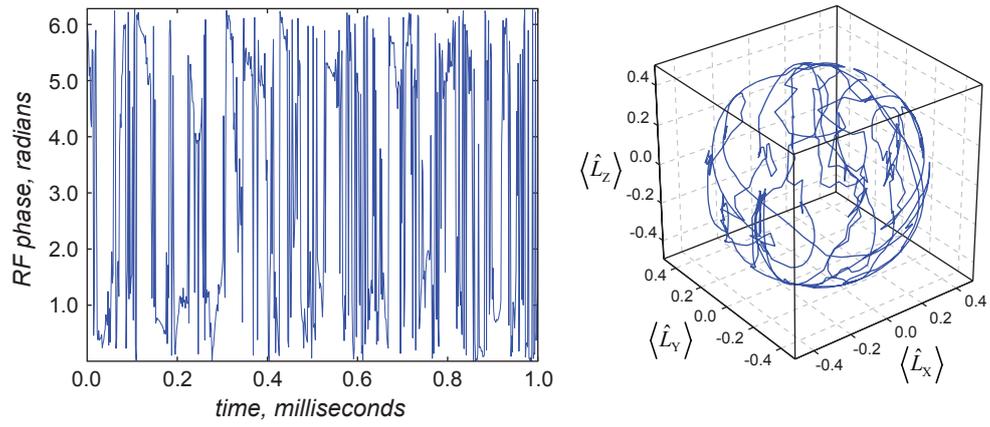

FIGURE 2

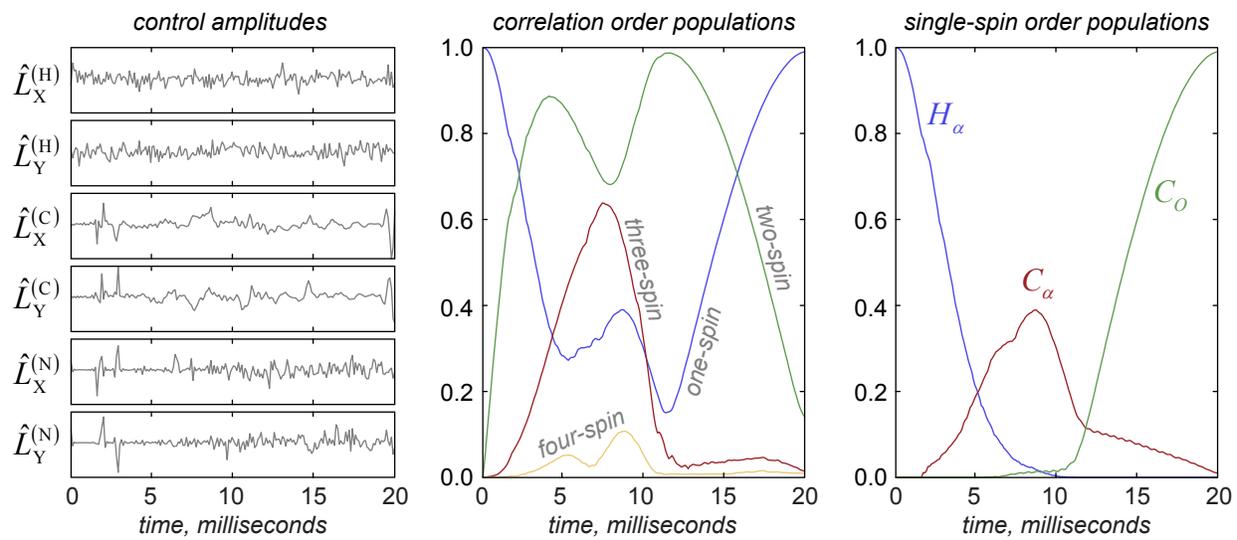

FIGURE 3

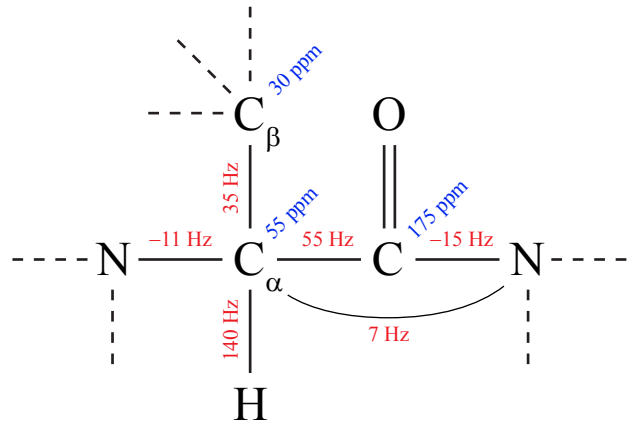

FIGURE 4

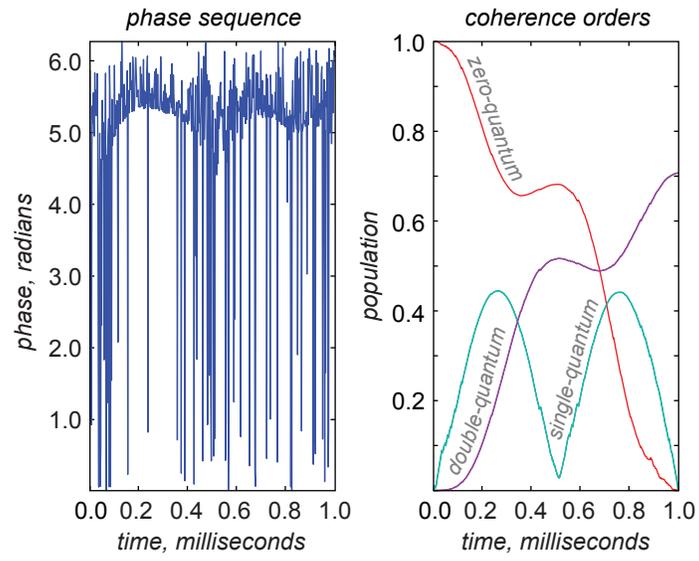

FIGURE 5

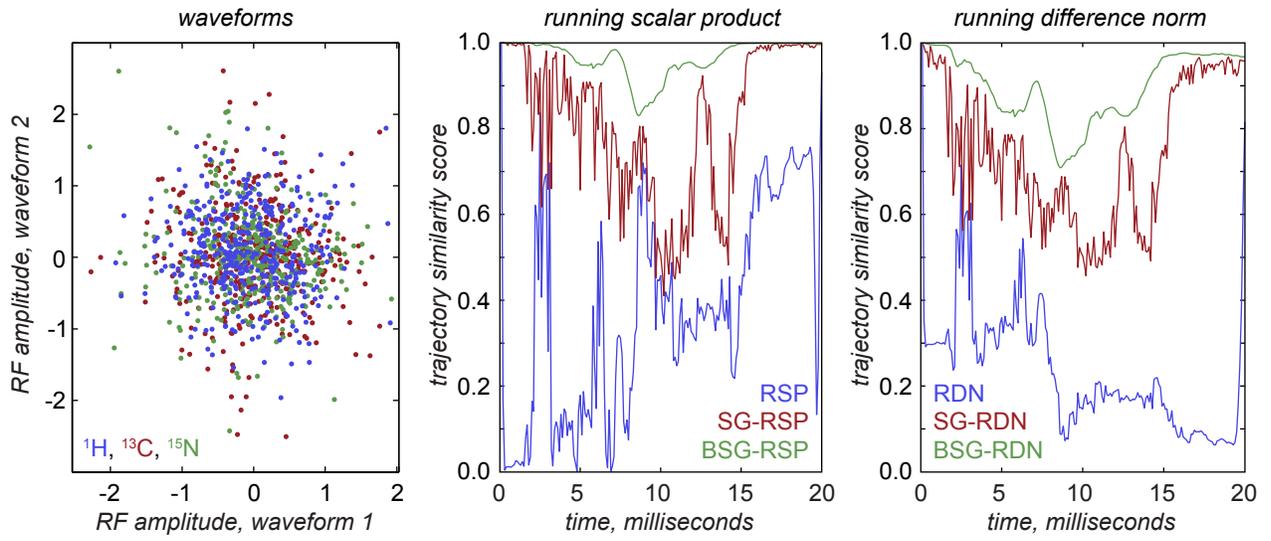